\documentclass[prc,preprint,showpacs,endfloats]{revtex4}
\usepackage{graphicx}
\usepackage{bm}
\usepackage{float}
\usepackage{amsmath}
\usepackage{epstopdf}
\usepackage{caption}

\begin{document}


\title{Derivation of the Grodzins relation in collective nuclear model}

\author{R.V. \surname{Jolos}}
\author{E.A. \surname{Kolganova}}
\affiliation{Joint Institute for Nuclear Research, 141980 Dubna, Russia}
\affiliation{Dubna State University, 141980 Dubna, Russia}

\begin{abstract}
Basing on the Bohr collective quadrupole Hamiltonian  the $A$-dependence of the Grodzins product is derived and the proportionality coefficient for the Grodzins relation is evaluated. The result obtained is in a correspondence with the experimental data.
\end{abstract}

\maketitle

In 1962 Grodzins \cite{Grodzins} analyzed the dependence of the E2 transition probabilities on the energy of the $0^+_1\rightarrow 2^+_1$ transitions. He has found for 126 E2 transition probabilities over the entire periodic table, known at that time, that the product $E(2^+_1)\times B(E2;0^+_1\rightarrow 2^+_1)$ vary as a smooth function of $A$ and $Z$. This property does not depend on whether the nucleus is spherical or deformed, although both $E(2^+_1)$ and
$B(E2;0^+_1\rightarrow 2^+_1)$  vary through a large factor. Later on Raman et al. \cite{Raman1,Raman2} analyzing a larger set of the experimental data have shown that the Grodzins relation can be presented in the following form
\begin{eqnarray}
\label{Eq1}
E(2^+_1)\times B(E2;0^+_1\rightarrow 2^+_1)=2.57(45)Z^2A^{-2/3}
\end{eqnarray}
where $E(2^+_1)$ is given in keV and $B(E2;0^+_1\rightarrow 2^+_1)$ in $e^2b^2$. This relation has been often used to estimate the unknown $B(E2;0^+_1\rightarrow 2^+_1)$ values from the known $E(2^+_1)$  in different nuclei, especially in nuclei close to the nuclear drip line. A generalization of the Grodzins relation for the case of the other E2 transitions within the ground state rotational band was considered in \cite{Jolos1,Gupta1}. Recently, Gupta has studied the global validity of the equation (\ref{Eq1}) and indicated the break down of the Grodzins product rule for some isotopes \cite{Gupta2,Gupta3}.  The value of the proportionality coefficient in the relation (\ref{Eq1}) has been considered in details in \cite{Pritychenko} using the latest set of the experimental data. The analysis indicates on a strong reason for individual fit of the proportionality coefficient in (\ref{Eq1}) for separate groups of nuclei. This means that the theoretical output of the proportionality coefficient in (\ref{Eq1}) obtained for all nuclei of the nuclide chart can only  qualitatively correspond to experimental data. As a consequence, it is important to find out the theoretical foundation of the Grodzins relation basing on one of the nuclear models.
Since both observables presented in the Grodzins relation characterize the collective nuclear properties, it is natural to choose a collective model as such a model.

The aim of the present paper is to derive an $A$-dependence of the Grodzins product and   qualitatively evaluate the value of the proportionality coefficient in (\ref{Eq1}) basing on the Bohr collective Hamiltonian.

Let's consider some consequences of the collective quadrupole model with Bohr  Hamiltonian. In the limiting case of the five-dimensional harmonic oscillator the E2 transition probability is
\begin{eqnarray}
\label{Eq2}
B(E2;0^+_1\rightarrow 2^+_1)=5\left(\frac{3}{4\pi}ZeR^2_0\right)^2\frac{\hbar}{2\sqrt{B_2C_2}},
\end{eqnarray}
and
\begin{eqnarray}
\label{Eq3}
E(2^+_1)=\hbar\sqrt{\frac{C_2}{B_2}}
\end{eqnarray}
Here, $C_2$ is the stiffness coefficient of the oscillator potential, $B_2$ is the inertia coefficient and $R_0=r_0A^{1/3}$ with
$r_0\approx 1.2\hspace{.1cm} {\rm fm}$.
Thus,
\begin{eqnarray}
\label{Eq4}
E(2^+_1)\times B(E2;0^+_1\rightarrow 2^+_1)=\frac{5}{2}\left(\frac{3}{4\pi}ZeR^2_0\right)^2\frac{\hbar^2}{B_2}
\end{eqnarray}

In the limit of the well deformed nuclei
\begin{eqnarray}
\label{Eq5}
B(E2;0^+_1\rightarrow 2^+_1)=\left(\frac{3}{4\pi}ZeR^2_0\right)^2\beta^2_2,
\end{eqnarray}
and
\begin{eqnarray}
\label{Eq6}
E(2^+_1)=\frac{3\hbar^2}{\Im},
\end{eqnarray}
where $\beta_2$ is the quadrupole deformation and $\Im$ is the moment of inertia. In the Bohr collective quadrupole Hamiltonian $\Im=3B_2\beta^2_2$. Then
\begin{eqnarray}
\label{Eq7}
E(2^+_1)\times B(E2;0^+_1\rightarrow 2^+_1)=\left(\frac{3}{4\pi}ZeR^2_0\right)^2\frac{\hbar^2}{B_2}.
\end{eqnarray}
Comparing (\ref{Eq4}) and (\ref{Eq7}) we see a discontinuity which is not observed experimentally.

From the relations (\ref{Eq4}) and (\ref{Eq7}) follows that the smooth variation of the Grodzins product with $A$ may be due to a smooth variation of $B_2$ with $A$. A break in transition from (\ref{Eq4}) to (\ref{Eq7}) indicates that proportionality coefficient in the Grodzins relation can depend on the character of the collective quadrupole motion in a nucleus.

The Grodzins product (\ref{Eq1}) looks as the term in the energy weighted sum rule and, therefore, can be obtained analyzing the double commutator of the quadrupole operator $Q_{2\mu}$ with the collective Bohr Hamiltonian.
The collective Hamiltonian has a form
\begin{eqnarray}
\label{Eq8}
H=-\frac{\hbar^2}{2B_2}\sum_{\mu}(-1)^{\mu}\frac{\partial^2}{\partial\alpha_{2\mu}\partial\alpha_{2-\mu}} +V(\alpha_{2\mu}).
\end{eqnarray}
In $H$ the potential $V$ can has very different form depending on  nucleus of which shape   is considered: spherical, transitional or deformed. In some cases potential  $V$ varies very strongly with $A$ and $Z$.  The inertia coefficient $B_2$ in general case depends on the collective variables $\alpha_{2\mu}$ \cite{Jolos2}. However, the main effect is contained in its deformation independent part. Thus, below we consider $B_2$ as a constant. In the collective model the electrical quadrupole moment operator is given by the expression  $Q_{2\mu}=\frac{3}{4\pi}ZeR^2_0\alpha_{2\mu}$. Then the double commutator looks as
\begin{eqnarray}
\label{Eq9}
[[H,Q_{2\mu}],Q_{2-\mu}]=-(-1)^{\mu}\frac{\hbar^2}{B_2}(\frac{3}{4\pi}ZeR^2_0)^2.
\end{eqnarray}
Taking the average of (\ref{Eq9}) over the ground state we obtain
\begin{eqnarray}
\label{Eq10}
\sum_n E(2^+_n)B(E2;0^+_1\rightarrow 2^+_n)=\frac{5}{2}\frac{\hbar^2}{B_2}(\frac{3}{4\pi}ZeR^2_0)^2,
\end{eqnarray}
where summation takes place over all collective quadrupole $2^+$ states related to the surface mode treated by the Bohr Hamiltonian.  Since E2 transition from the ground to the first $2^+$ state exceed significantly a contribution of the other $2^+$ states we can restrict summation in (\ref{Eq10}) by the first term. Keeping in (\ref{Eq10}) only $2^+_1$ state we obtain on the left hand side of (\ref{Eq10}) the Grodzins product and on the right hand side the expression containing quadrupole inertia coefficient $B_2$
\begin{eqnarray}
\label{Eq11}
E(2^+_1)\times B(E2;0^+_1\rightarrow 2^+_1)=\frac{5}{2}\frac{\hbar^2}{B_2}(\frac{3}{4\pi}ZeR^2_0)^2.
\end{eqnarray}
However, this approximation means that the expression on the right in (\ref{Eq11}) is the upper boundary for this coefficient.

It is interesting to mention that if we substitute in (\ref{Eq11}) the liquid drop expression for the inertia coefficient $B_2=3/(8\pi) AmR^2_0$ we obtain
\begin{eqnarray}
\label{Eq12}
E(2^+_1)\times B(E2;0^+_1\rightarrow 2^+_1)=\frac{15}{4\pi}\frac{\hbar^2 r_0^2}{m}(e Z)^2A^{-1/3}.
\end{eqnarray}
We see that the $A$-dependence of the Grodzins product given in (\ref{Eq12}) contradicts to that following from the analysis of the experimental data.

In the crancking model approach \cite{Moszkowski}  the following expression is derived for the inertia coefficient of the quadrupole mode
\begin{eqnarray}
\label{Eq13}
B_2=2\hbar^2\sum_k\frac{|\langle k|\frac{\partial}{\partial\alpha_{2\mu}}|g.s.\rangle |^2}{E_k-E_{g.s.}},
\end{eqnarray}
where the summation is performed over the excited states. Using the relation
\begin{eqnarray}
\label{Eq14}
\langle k |[H,\frac{\partial}{\partial\alpha_{2\mu}}]|g.s.\rangle=(E_k-E_{g.s.})\langle k |\frac{\partial}{\partial\alpha_{2\mu}}|g.s.\rangle .
\end{eqnarray}
and restricting summation in (\ref{Eq14})  by the single particle states of the shell model Hamiltonian with the deformed Woods-Saxon mean field potential we obtain
\begin{eqnarray}
\label{Eq15}
B_2=2\hbar^2\sum_{s,t}\frac{(n_s-n_t)|\langle s |rdV_{WS}/dr\frac{1}{\sqrt{2(1+\delta_{\mu 0})}}(Y_{2\mu}+Y_{2-\mu})|t\rangle |^2}{(E_s-E_t)^3},
\end{eqnarray}
where $s$ and $t$ are the single particle quantum numbers.
If we take into account  dependence of the coefficients of inertia on deformation, we would have to distinguish the coefficient for $\beta$-mode ($\mu=0$), $\gamma$-mode ($\mu=2$) and rotational motion
($\mu=1$) \cite{Jolos2}. However, we will neglect this dependence below and put $\mu=0$ for shortness of notations.

If we take into account the effect of pair correlations we obtain from (\ref{Eq15})
\begin{eqnarray}
\label{Eq16}
B_2=2\hbar^2\sum_{s,t}\frac{|\langle s |rdV_{WS}/dr Y_{20}|t\rangle |^2 (\varepsilon_s \varepsilon_t-(E_s-\lambda)(E_t-\lambda)+\Delta^2)}{2\varepsilon_s\varepsilon_t (\varepsilon_s+\varepsilon_t)^3},
\end{eqnarray}
where $\varepsilon_s=\sqrt{(E_s-\lambda)^2+\Delta^2}$ is the single quasiparticle energy.

The method of a qualitative consideration of the expression for $B_2$ is presented below.
Just a qualitative consideration makes it possible to determine  dependence of $B_2$ on $A$. This becomes possible due to using the average expressions for the pairing gap and density of single particle levels.
At the same time, however, the shell effects are far from fully be taken into account.

Since we are interested in application of the results obtained to heavy nuclei we
consider, as an example, the matrix elements $|\langle s|r dV_{WS}/dr Y_{20}|t\rangle|$
for two nuclei representing well deformed nuclei from the rare earth and actinide regions.
We have found that as a rule the diagonal matrix elements $|\langle s|r dV_{WS}/dr Y_{20}|s\rangle|$
are significantly larger than the nondiagonal ones. This means that in the sum (\ref{Eq16})
we can keep only terms with diagonal matrix elements.

Let us compare the absolute values of the matrix elements $|\langle s|r dV_{WS}/dr Y_{20}|s\rangle|$ obtained for both considered nuclei
averaged over the single particle levels near the Fermi surface in the energy interval
$\lambda \pm 5$ MeV. Here $\lambda$ is the chemical potential.
Just single particle levels near the Fermi surface give the main contribution into the sum (\ref{Eq16}).
As the result we obtained that the average values of the matrix elements practically coincide
for both considered nuclei: 10.5 for $^{168}$Er and 10.9 for $^{238}$U. Thus, we can consider these matrix elements as $A$-independent.
At least, their ratio
is closer to one than $\left(A_2(=238)/A_1(=168)\right)^{1/3}$. The last result characterizes the accuracy
with which the $A$-dependence of the right side of the Grodzins relation is obtained below.

Taking only diagonal matrix
elements in (\ref{Eq16}) and introducing the average matrix element instead of the single particle
state dependent ones we obtain
\begin{eqnarray}
\label{Eq17}
B_2=2\hbar^2|\langle s|r dV_{WS}/dr Y_{20}|s\rangle|^2_{average}\times\sum_s\frac{\Delta^2}{8\epsilon_s^5}
\end{eqnarray}
We can estimate the sum over the single particle states in (\ref{Eq17}) integrating instead of summing.
Clearly, the shell effects are neglected in this case.
However, it gives a possibility to determine an $A$-dependence of the whole expression which is one of our aims.
Thus,
\begin{eqnarray}
\label{Eq18}
\sum_s\frac{\Delta^2}{8\varepsilon_s^5}\rightarrow \frac{1}{8}\int gd(E_s-\lambda)\frac{\Delta^2}{\left((E_s-\lambda)^2+\Delta^2\right)^{5/2}}=\frac{g}{6\Delta^2}.
\end{eqnarray}
Here $g$ is the single particle level density near the Fermi level. We obtain this result if we do not limit the integration interval.
If we limit the integration interval of $|E_s-\lambda|$ by 2$\Delta$, then the factor on the right hand side of (\ref{Eq18}) will be only 2$\%$ less.
Substituting (\ref{Eq18})  in (\ref{Eq17}) we obtain the approximate expression for $B_2$.
Substituting this result into
(\ref{Eq11}) and using the average value of the energy gap $\Delta=12/\sqrt{A}$ MeV \cite{BM1},
the calculated average value of the matrix element $|\langle s|r dV_{WS}/dr Y_{20}|s\rangle|$
 and the Fermi gas single particle level density parameter value near the Fermi
level $g=3A/2\epsilon_F$ \cite{BM1}, where $\epsilon_F$ is the Fermi energy,
we obtain for the Grodzins product the following result
\begin{eqnarray}
\label{Eq19}
E(2^+_1)\times B(E2;0^+_1\rightarrow 2^+_1)=2.9\frac{Z^2}{A^{2/3}} (eb)^2{\rm keV}
\end{eqnarray}
We should not overestimate the meaning of the fact of a practically coincidence of the calculated
proportionality coefficient with the phenomenological one since the calculations are performed only
qualitatively. Rather, it indicates that the approach used to derive the Grodzince
relation is reasonable.

In conclusion, we derived the $A$-dependence of the Grodzins product basing on the Bohr collective quadrupole Hamiltonian and qualitatively evaluate the proportionality coefficient for the Grodzins relation. The resulting values presented in  (\ref{Eq19}) is in an  agreement with the experimental value.

In the next paper we are planning to perform the microscopic calculations of $B_2$  for the superheavy nuclei and basing on these calculations make predictions for the energies of the $2^+_1$ states in these nuclei using the calculated values of the quadrupole deformation.

\bigskip

Authors acknowledge the partial support from the Heisenberg--Landau Program. RVJ acknowledge support by the Ministry of Education and Science (Russia) under Grant No. 075-10-2020-117. EAK acknowledge support by the Russian Foundation for Basic Research under Grant No. 20-02-00176.


\begin{thebibliography}{99}
\bibitem{Grodzins} L.~Grodzins,  Phys.Lett. B {\bf 2}, 88 (1962).
\bibitem{Raman1} S.~Raman, C.W.~Nestor, T.~Tikkanen, At. Data Nucl. Data Tables {\bf 78}, 1 (2001).
\bibitem{Raman2} S.~Raman, C.W.~Nestor, Jr., K.H.~Bhatt, Phys.Rev. C {\bf 37}, 805 (1988).
\bibitem{Jolos1} R.V.~Jolos, P.~von Brentano, N.~Pietralla, Phys. Rev. C {\bf 71}, 044305 (2005).
\bibitem{Gupta1} J.B.~Gupta and V.~Katoch, Int.J.Mod.Phys. E {\bf 28}, 1950055 (2019).
\bibitem{Gupta2} J.B.~Gupta, Phys. Rev. C {\bf 89}, 034321 (2014).
\bibitem{Gupta3} J.B.~Gupta, Nucl. Phys. A, {\bf 978}, 25 (2018).
\bibitem{Pritychenko} B.~Pritychenko, M.~Birch, B.~Singh, Nucl. Phys. A {\bf 962}, 73 (2017).
\bibitem{Jolos2} R.V.~Jolos and P.~von Brentano, Phys. Rev. C {\bf 79}, 044310 (2009).
\bibitem{Moszkowski} S.A.~Moszkowski, in {\it Encyclopedia of Physics, Structure of Atomic Nuclei, ed. by S.~Fl\"ugge,}, Vol. XXXIV.
\bibitem{BM1} A.~Bohr and B.R.~Mottelson, {\it Nuclear Structure}, vol. I (Benjamin, New York, 1969).





\end{thebibliography}
\end{document}